\documentclass[10pt,twocolumn]{article} 

 \usepackage{multirow}

\usepackage{simpleConference}
\usepackage{times}
\usepackage{booktabs}       
\usepackage{graphicx}
\usepackage{amssymb}
\usepackage{url}
\usepackage[hidelinks]{hyperref}

\usepackage{subcaption}

\DeclareCaptionFormat{custom}
{%
    #1 #2\textit{\small #3}
}
\begin{document}

\title{Neural TTS in French: Comparing Graphemic and Phonetic Inputs Using the SynPaFlex-Corpus and Tacotron2}

\author{Samuel Delalez, Ludi Akue \\
\\
Lunii\\
\texttt{\{samuel, ludi\}@lunii.com}  \\
}

\maketitle
\thispagestyle{empty}

\begin{abstract}

The SynPaFlex-Corpus is a publicly available TTS-oriented dataset, which provides phonetic transcriptions automatically produced by the JTrans transcriber, with a Phoneme Error Rate (PER) of 6.1\%.
In this paper, we analyze two mono-speaker Tacotron2 models trained on graphemic and phonetic inputs, provided by the SynPaFlex-Corpus. Through three subjective listening tests, we compare their pronunciation accuracy, sound quality and naturalness. Our results show significantly better pronunciation accuracy and prosody naturalness for the phoneme-based model, but no significant difference in terms of perceived sound quality. They demonstrate that a PER of 6.1\% is sufficient to enhance pronunciation control by using phonetic transcripts instead of graphemes with 83 hours of recorded French read speech. They suggest that the SynPaFlex-Corpus is suitable for pre-training a model in mono-speaker fine-tuning approaches.
\end{abstract}

\keywords{Speech Synthesis \and Tacotron \and Phonemes \and Dataset \and SynPaFlex-Corpus \and Neural Text-To-Speech}

\section{Introduction}
\label{sec:intro}

Neural speech synthesizers are widely used for their ability to realistically imitate human voices, but the amount of data required to train modern zero-shot Text-To-Speech (TTS) systems \cite{jia2018transfer, tang2021zero, wang2023neural} may not be available for all languages or organizations. Pre-training and fine-tuning conventional cascaded TTS systems \cite{shen2018natural, ren2020fastspeech, li2019neural} is still a relevant use-case \cite{bollepalli2019lombard, tits2019exploring, bird2019phoneme, fahmy2020transfer, gopalakrishnan2022fine}. Yet, because these systems are trained on relatively small amounts of data, they still suffer from issues with pronunciation accuracy \cite{perquin2020investigation, taylor2021liaison, taylor2022pronunciation}, especially when synthesizing proper nouns or loan words. It is crucial for pre-trained models to be as accurate as possible in order to prevent fine-tuned models from inheriting pronunciation inaccuracies.

Graphemic representations of languages with non-phonetic orthography, such as French (the language of interest in this paper), can have ambiguous pronunciation, which sometimes strongly depends on context (e.g. \emph{fils} [fil]: french for \emph{threads}; \emph{fils} [fis]: french for \emph{son}). Using phonemes as inputs instead of graphemes might improve pronunciation control \cite{jia2018transfer, bird2019phoneme,taylor2022pronunciation, fong2019comparison}, as correct phonetic transcriptions remove any pronunciation ambiguity. 

Existing work \cite{perquin2020investigation}, however, found no significant difference in terms of pronunciation and sound quality between two Tacotron2 models trained respectively on grapheme and phoneme inputs, using the SIWIS dataset \cite{honnet2017siwis}, with automatic phonetic transcriptions performed by the e-speak synthesizer\footnote{\url{https://espeak.sourceforge.net/}}.

Following this work, we compared two Tacotron2 models,  trained on graphemic and phonetic transcriptions provided by the SynPaFlex-Corpus \cite{sini2018synpaflex}. The automatic phonetic transcriptions of this dataset were performed by the JTrans phonetizer \cite{cerisara2009jtrans}, with an average PER of 6.1\%. Our main objective is to verify whether this PER is sufficiently small to enhance pronunciation accuracy by using the phonetic transcripts instead of graphemes. We also aim to compare perceived sound quality and prosody naturalness. 

We conducted three subjective listening tests, presented in Section \ref{sec:XP}. The results, discussed in Section \ref{sec:discussion}, showed significantly better pronunciation accuracy and prosody naturalness for the phonetic model, but no significant difference in terms of perceived sound quality.
 
\section{Experiments}
\label{sec:XP}

Our experiments were designed to compare three aspects between our grapheme-based and phoneme-based TTS models.

\begin{itemize}
    \item \textbf{Perceived pronunciation accuracy}: the ability for a model to produce a speech utterance with the expected pronunciation by listeners when compared to a reference text.
    \item \textbf{Overall sound quality}: speech signals with less synthesis artifacts should be judged of better quality.
    \item \textbf{Prosody naturalness}: synthesized speech rhythm and intonation is perceived as natural when it convincingly sounds like if a human could have uttered the sentence.
\end{itemize}

\subsection{TTS models}
\label{sec:TTS_models}
All of our models were trained on a single NVIDIA Tesla V100 GPU with 32GB of memory. We used the implementations provided in the publicly available TensorFlowTTS\footnote{\url{https://github.com/TensorSpeech/TensorFlowTTS}} repository.

Training setup is summarized in Table \ref{tab:training_setup}. For mel-spectrogram generation, we trained two distinct Tacotron2 models: one model with graphemic inputs, noted as \texttt{Graph}, and the other with phonetic inputs, noted as \texttt{Phon}. Both models were trained for 200k steps. Most of the overall Tacotron2 architecture and training setup were kept identical to the original paper \cite{shen2018natural}, except for the number of convolutional layers in the encoder set to 5 and batch size to 32.
We used Multi-Band MelGAN (MB-MelGAN) \cite{yang2021multi} as a neural vocoder and trained a single model for 500k steps. The initial learning rate was set to $2.5e-3$, and the batch size to 64. The rest was kept identical to the original paper.

\begin{table}[]
\centering
\caption{Training setup for Tacotron2 \emph{\cite{shen2018natural}} and MB-MelGAN \emph{\cite{yang2021multi}}. If not specified, parameters were kept identical to the original papers.}
\begin{tabular}{l|ll}
\midrule
\multirow{3}{*}{Tacotron2} & Training steps        & 200k     \\
                           & Encoder conv layers   & 5        \\
                           & Batch size            & 32       \\
                           \midrule

\multirow{3}{*}{MB-MelGAN} & Training steps        & 500k     \\
                           & Initial learning rate & $2.5e-5$ \\
                           & Batch size            & 64      \\
                           \midrule

\end{tabular}
\label{tab:training_setup}
\end{table}

\subsection{Training data}
\label{sec:synpaflex}

 The SynPaFlex-Corpus \cite{sini2018synpaflex} is a publicly available TTS-oriented dataset that contains 87 hours of French read speech, including text and audio recorded by a single female speaker. It has been extracted from the Librivox library and reorganized by the authors. They performed automatic phonetic transcription and forced alignment with JTrans \cite{cerisara2009jtrans}, and stated that when compared to human annotations, the average Phoneme Error Rate (PER) is 6.1\%. Other automatic data preparation included loudness harmonization, text format unification, and text normalization.
Every audio example was downsampled to 22.05kHz. To prevent Out Of Memory issues, audio examples must not exceed 12 seconds according to our training setup. We used the author's forced alignment results to split the dataset into [text and phonemes]/audio pairs that fulfill this requirement. We extracted 5\% of the dataset as validation data, leading to around 83 hours of training data.

\subsection{Experimental material}

We selected 100 sentences from our validation set that are as difficult as possible to pronounce. We prioritized consonant and vowel cluster diversity in our selection process, resulting in sentences that include 9 distinct consonant quadriphones, 76 distinct consonant triphones, and 39 distinct vowel diphones.
Each sentence was synthesized with both our models, and the ground truth mel-spectrograms were converted to audio signals using MB-MelGAN \cite{yang2021multi} and the Griffin-Lim algorithm \cite{griffin1984signal}.

\subsection{Phoneme Error Rate (PER)}
\label{sec:PER}

\begin{table}[]
\centering
\caption{PER estimated by two experts for \emph{\texttt{Graph}} and \emph{\texttt{Phon}}.}
\begin{tabular}{lll}
\midrule
                          & Model & PER (\%) \\
                           \midrule

\multirow{2}{*}{Expert 1} & \texttt{Graph} & 1.48     \\
                          & \texttt{Phon}  & 0.39     \\
                           \midrule

\multirow{2}{*}{Expert 2} & \texttt{Graph} & 1.5         \\
                          & \texttt{Phon}  & 0.53         \\
                           \midrule
\multirow{2}{*}{Average}  & \texttt{Graph} & 1.49         \\
                          & \texttt{Phon}  & 0.46       \\
                          \midrule
\end{tabular}
\label{tab:PER}
\end{table}

In order to estimate the PER of our models, two experts were asked to listen to all test sentences synthesized by both models (around 19 minutes of audio per model), and to quantify the number of phoneme errors for each of them. Phoneme errors included substitutions, deletions, and disallowed \emph{liaisons} (a French habit of pronunciation which consists in uniting the last consonant of a word usually not pronounced with the initial vowel of the following word, e.g. \emph{me\underline{s} amis} [me\underline{z}ami]) \footnote{\href{https://www.larousse.fr/dictionnaires/francais/liaison/46939}{Definition extracted from the Larousse dictionnary}}. 

Results are presented in Table \ref{tab:PER}, showing better performances for \texttt{Phon}, with an average PER of $0.46\%$. This PER is much lower than the PER of $6.1\%$ of our training data. This indicates that the amount of data used for training our phoneme-based model allows for smoothing the errors produced by the JTrans phonetizer.

\subsection{Listening tests}

\begin{table}[]
\centering
\caption{ANOVA results for all pairs of conditions in our pronunciation evaluation.}
\begin{tabular}{ll}
\midrule
\multicolumn{1}{l}{} & $p$ (ANOVA) \\
\midrule
\texttt{GT}$_{22}$ / \texttt{V-GT}    & $3.2e-3$               \\
\texttt{V-GT} / \texttt{Phon}     & $0.03$                \\
\texttt{V-GT} / \texttt{Graph}     & $3.5e-8$            \\
\texttt{GT}$_{22}$ / \texttt{Phon}      & $3.2e-6$           \\
\texttt{GT}$_{22}$ / \texttt{Graph}      & $2.6e-14$           \\
\texttt{Graph} / \texttt{Phon}       & $6.9e-4$       \\
\midrule
\end{tabular}
\label{tab:ANOVA}

\end{table}

\begin{figure}[]
  \centering
  \includegraphics[width = 0.48\textwidth]{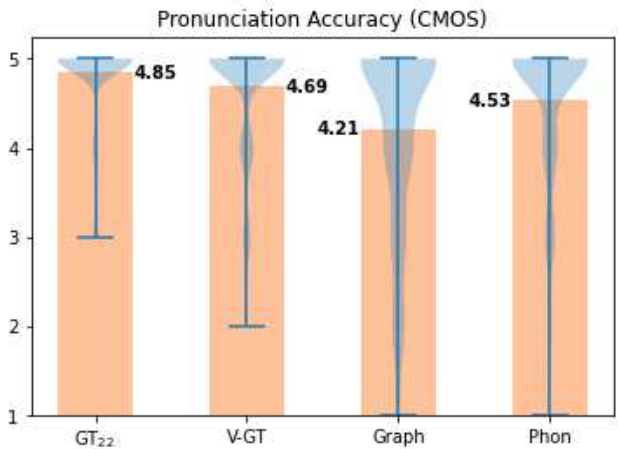}
  \caption{CMOS pronunciation evaluation results. Orange bars represent mean scores. The exact mean score values are displayed next to each bar. Blue violin plots represent scores distribution. Horizontal blue lines represent maximal and minimal scores.}
  \label{fig:MOS_pronunciation}
\end{figure}

We followed the recommendations of \cite{wester2015we} for conducting our experiments. 42 native French participants from different backgrounds, 19\% of which are audio books experts as they work in audio book publishing, took three distinct listening tests in order to assess pronunciation accuracy, overall sound quality and prosody naturalness of our models. 
The graphical interface was developed using the WebMUSHRA framework \cite{schoeffler2018webmushra}. We made sure that each participant was presented different sentences for each test. All 100 sentences were judged by at least two participants for each test.

\subsubsection{Perceived pronunciation accuracy}
\label{sec:pronunciation}

We evaluated pronunciation accuracy in a CMOS test. We randomly selected 5 sentences for each participant. For each sentence, we presented 4 samples on a webpage: ground truth sampled at 22.05kHz, noted as \texttt{GT}$_{22}$, vocoded ground truth using MB-MelGAN, noted as \texttt{V-GT}, synthesized signals from our grapheme-based model, noted as \texttt{Graph}, and from our phoneme-based model, noted as \texttt{Phon}. Participants were asked to rate pronunciation from 1 to 5 by comparing the audio examples with a reference text, with the following instructions: 1 - The audio does not match the text; 2 - Errors are very annoying; 3 - Errors are slightly annoying; 4 - Errors are perceptible but not annoying; 5 - No pronunciation error. We specifically asked participants not to judge sound quality and to ignore unnatural prosody.

The results are presented in Figure \ref{fig:MOS_pronunciation}, and Table \ref{tab:ANOVA} shows the $p$ values obtained with ANOVA for every pair of conditions. Results are statistically significant if $p < 0.05$. It is interesting to note that \texttt{GT}$_{22}$ is significantly higher rated than \texttt{V-GT}, even though they contained the exact same utterances. This indicates that sound quality degradation introduced by the vocoder might have been judged as pronunciation errors. However, \texttt{Phon} is significantly higher rated than \texttt{Graph}. These results are consistent with the lower PER estimation for \texttt{Phon} observed in Section \ref{sec:PER}. This indicates that the JTrans automatic phonetic transcriber is precise enough to achieve better pronunciation performances by using phonemes rather than graphemes with the amount of data we used. 

Our observations differ from those of \cite{perquin2020investigation}, where pronunciation was perceived as equivalent for both models. This discrepancy may be due to differences in the automatic phonetic transcriber (JTrans vs e-speak) or the amount of data (83 hours vs 10 hours) used.

\subsubsection{Overall sound quality}

In a second test, we assessed the overall sound quality using a MUSHRA-like test. For each participant, 5 new sentences were randomly selected. For each sentence, 5 examples were presented. In addition to \texttt{Graph}, \texttt{Phon}, and \texttt{V-GT} (see Section \ref{sec:pronunciation}) we used high and low-range anchors: respectively, the ground truth signal sampled at 44.1kHz, noted as \texttt{GT}$_{44}$, and the ground truth mel-spectrogram converted into audio via the Griffin-Lim algorithm, noted as \texttt{GL-GT}. No reference signal was presented. Participants were asked to rate the overall sound quality between 0 and 100, with the following instructions: Mediocre (0-20), Bad (20-40), Tolerable (40-60), Good (60-80) and Excellent (80-100). We specifically asked participants to ignore mispronunciations or unnatural prosody. No reference text was presented.

\begin{figure}
  \centering
  \includegraphics[width = 0.48\textwidth]{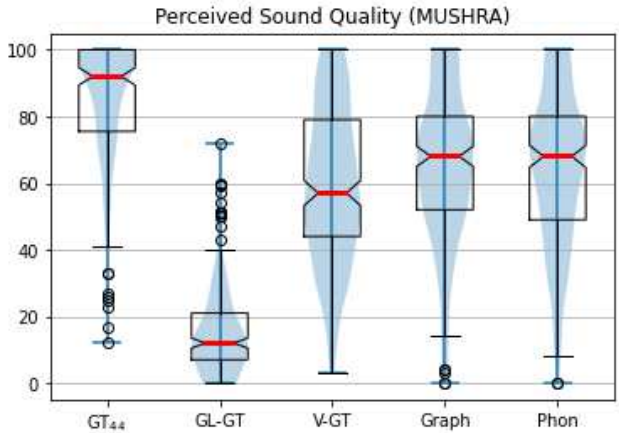}
  \caption{Perceived overall sound quality MUSHRA results. Red lines represent median values. Boxplots represent interquartile range, and notches 95\% confidence interval. Circles represent outliers, Horizontal black lines represent minimal and maximal scores, excluding outliers. Violin plots represent scores distribution. Horizontal blue lines represent maximal and minimal scores.}
  \label{fig:MUSHRA_sound_quality}
\end{figure}

The results are presented in Figure \ref{fig:MUSHRA_sound_quality}. Unsurprisingly, our high-range anchor \texttt{GT}$_{44}$ is rated as "Excellent", and our low-range anchor \texttt{GL-GT} as "Mediocre". \texttt{Graph} and \texttt{Phon} were rated around 70 with no significant difference, which corresponds to a "Good" quality. Surprisingly, \texttt{V-GT} was rated significantly worst than \texttt{Graph} and \texttt{Phon}, and under 60, which corresponds to a "Tolerable" Quality. 
We hypothesize that mel-spectrogram conversion of ground truth utterances with more prominent pitch and timbre variations than synthesized utterances might be more challenging for the vocoder. It would be interesting to verify if additional MB-MelGAN training (e.g. 1M steps) would reduce this gap. 

Our results indicate that sound quality is not deteriorated by Tacotron2 synthesis when compared to \texttt{V-GT}, and that using graphemic or phonetic inputs is equivalent in terms of sound quality, which supports the findings of \cite{perquin2020investigation}.

\subsubsection{Prosody naturalness}

In a third test, we compared the prosody naturalness performance via a paired preference test. We randomly selected 5 new sentences for each participant. For each sentence, we presented two examples: one produced by \texttt{Graph} and the other by \texttt{Phon}. Participants were asked to select which example sounded the most natural to them in terms of prosody, and we specified that sound quality should be ignored.

\begin{figure}
  \centering
  \includegraphics[width = 0.38\textwidth]{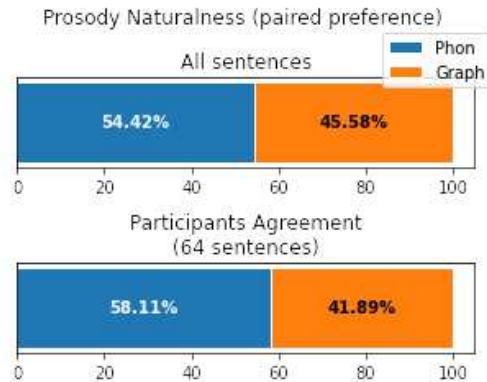}
  \caption{Naturalness paired preference results. Top: The results include all 100 test sentences. Bottom: The results only include the 64 sentences that received the same preference by at least two participants.}
  \label{fig:A:B_naturalness}
\end{figure}

The results are presented in Figure \ref{fig:A:B_naturalness}. When analyzing all 100 test sentences (top of Figure \ref{fig:A:B_naturalness}), we found a slight preference for \texttt{Phon}, but the results are not significant ($p = 0.067$). 
Since the task was judged as difficult by the participants, we excluded the sentences that the participants did not agree on. Out of the 100 sentences, 64 sentences received the same naturalness preference by at least two participants. In this situation \texttt{Phon} is significantly preferred (bottom of Figure \ref{fig:A:B_naturalness}, $p = 0.012$).

\section{Discussion and perspectives}
\label{sec:discussion}

The results presented in Section \ref{sec:XP} demonstrate that using phonetic transcriptions provided in the SynPaFlex-Corpus produces better results than using corresponding graphemic representations when training a Tacotron2 model. When compared to our grapheme-based model, our phoneme-based model shows a lower estimated PER, better perceived pronunciation accuracy and prosody naturalness while maintaining equivalent sound quality.

Our results do not corroborate the findings of \cite{perquin2020investigation}, where perceived pronunciation accuracy was equivalent with both input types. It is important to note that the authors did not use the same amount of data. The SIWIS dataset only contains 10 hours of read speech, and they did not use the same phonetic transcriber. Therefore, it would be interesting to test 1) whether our results still hold if we train our models with only 10 hours of speech from the SynPaFlex-Corpus, and 2) whether replacing the e-speak transcriber with JTrans for automatic phonetic transcriptions of the SIWIS dataset would allow for phonetic inputs to improve pronunciation accuracy.

Our results indicate that the PER of the JTrans transcriber is small enough to improve pronunciation accuracy when training Tacotron2 with phonemes instead of graphemes, with 83 hours of speech. 
Our phoneme-based model has a lower estimated PER than the JTrans automatic phonetic transcriptions provided in the SynPaFlex-Corpus.
 This suggests that phonetic errors are statistically smoothed out by training Tactoron2 with this amount of data. 

The SynPaFlex-Corpus appears to be suitable for pre-training a model for further fine-tuning. Future work will evaluate this hypothesis and determine the minimum amount of data for our observations to be replicated with smaller datasets in a mono-speaker fine-tuning approach.

\section{Acknowledgements}
This work was partially supported by the French Agence Nationale de la Recherche (ANR), under grant ANR-21-CE23-0040 (project EXOVOICES). 
This work was granted access to the HPC resources of IDRIS under the allocation 2022-AD011012954R1 made by GENCI.

\bibliographystyle{ieeetr}
\bibliography{main}
\end{document}